\documentstyle[12pt]{article}

\begin{document}
\setcounter{page}{1}
\renewcommand{\thefootnote}{\fnsymbol{footnote}}
\pagestyle{plain} \vspace{1cm}
\begin{center}
\Large{\bf Gravitational Uncertainty and Black Hole Remnants }\\
\small \vspace{1cm} {\bf Kourosh Nozari}{\footnote{
knozari@umz.ac.ir}}\quad\quad and \quad\quad {\bf S. Hamid
Mehdipour}{\footnote{h.mehdipour@umz.ac.ir}} \\
\vspace{0.5cm} {\it Department of Physics, Faculty of Basic
Sciences,
University of Mazandaran,\\
P. O. Box 47416-1467, Babolsar, IRAN}\\
October 2005\\

\end{center}
\vspace{1.5cm}

\begin{abstract}
Possible existence of black holes remnants provides a suitable
candidates for dark matter. In this paper we study the possibility
of existence for such remnants. We consider quantum gravitational
induced corrections of black hole's entropy and temperature to
investigate the possibility of such relics. Observational scheme for
detection of these remnants and their cosmological
constraints are discussed.
\end{abstract}
\newpage

\section{Introduction}
It is by now widely accepted that dark matter (DM) constitutes a
substantial fraction of the present critical energy density in the
Universe. However, the nature of DM remains an open problem. There
exist many DM candidates, most of them are non-baryonic weakly
interacting massive particles (WIMPs), or WIMP-like particles [1].
By far the DM candidates that have been more intensively studied are
the lightest supersymmetric (SUSY) particles such as neutralinos or
gravitinos, and the axions (as well as the axinos). There are
additional particle physics inspired dark matter candidates [1]. A
candidate which is not as closely related to particle physics is the
relics of primordial black holes (Micro Black Holes) [2,3]. Certain
inflation models naturally induce a large number of such a black
holes. As a specific example, hybrid inflation can in principle
yield the necessary abundance of primordial black hole remnants for
them to be the primary source of dark matter [4,5]. In recent years
it has been suggested that measurements in quantum gravity should be
governed by generalized uncertainty principle (GUP). In fact, some
evidences from string theory, quantum geometry and black hole
physics, have led some authors to re-examine usual uncertainty
principle of Heisenberg [6-13]. These evidences have origin on the
quantum fluctuations of the background space-time metric. Existence
of a minimal length scale on the order of Planck length is an
immediate consequence of GUP. Introduction of this idea has attract
considerable attention and many authors considered various problems
in the framework of generalized uncertainty principle [14-28]. The
issue of black holes remnants has been considered by some authors.
Adler and his coworkers have argued that contrary to standard
viewpoint, GUP may prevent small black holes total evaporation in
exactly the same manner that the uncertainty principle prevents the
Hydrogen atom from total collapse [29]. Chen considering inflation
induced primordial black holes, have investigated the issue of
stability of such relics [30]. Recently, Varying Speed of Light
(VSL), as a new conjecture, which has been proposed to solve the
problems of standard cosmology, has attract some attentions. After
introduction of this conjecture, several alternative VSL theories
have been proposed and some of their novel implications have been examined extensively [31-35].\\
In this paper the issue of black hole remnants will be considered in
the framework of both GUP and VSL. The main consequence of this
combination is related to the stability problem of possible
remnants. In forthcoming sections first we show that GUP provides a
reasonable framework for VSL. Then using a simple VSL model we will
show that if one consider both the effect of GUP and VSL on the
thermodynamics of black holes, the results of Bekenstein-Hawking
concerning total evaporation of black holes should be re-examined.
In Bekenstein-Hawking approach the total evaporation of micro black
hole is possible. Here we will see that it is possible to have
relics of evaporating black holes which can be considered as a
possible candidate for dark matter. The structure of the paper is as
follows: in section 2 we show that VSL can be considered as an
immediate consequence of GUP. Section 3 is devoted to the quantum
gravitational corrected black hole thermodynamics. Some numerical
calculations have been down and their physical results are
discussed. The paper follows by conclusions and discussion regarding
observational scheme for detection of such relics and their
cosmological constraints in section 4.

\section{Preliminaries}
As has been revealed in introduction, usual uncertainty principle of
quantum mechanics, the so-called Heisenberg uncertainty principle,
should be re-formulated due to noncommutative nature of spacetime at
Planck scale. As a consequence, it has been indicated that in
quantum gravity there exists a minimal observable distance on the
order of the Planck length which governs on all measurements in
extreme quantum gravity limit. In the context of string theories,
this observable distance is referred to GUP. A generalized
uncertainty principle can be formulated as
\begin{equation}
\Delta x\geq \frac{\hbar}{\Delta p} + const. G\Delta p,
\end{equation}
which, using the minimal nature of $l_P$ can be written as,
\begin{equation}
\Delta x\geq\frac{\hbar}{\Delta p}+\frac{\alpha^{\prime} l_p^2\Delta
p}{\hbar}.
\end{equation}
The main consequence of GUP is that measurement of the position is
possible only up to Planck length, $l_{P}$. So one can not setup a
measurement to find more accurate particle position than Planck
length, and this means that the notion of locality breaks down. It
is important to note that there are more generalization which
contain further terms in right hand side of equation (2) (see [36]),
but in some sense regarding dynamics, equation (2) has more powerful
physical grounds. Suppose that
$$\Delta x\sim x,\quad\quad\Delta p\sim p,\quad\quad p=\hbar k,\quad\quad x=\bar{\lambda}=\frac{\lambda}{2\pi}.$$
Therefore one can write,
\begin{equation}
\bar{\lambda}=\frac{1}{k}+\alpha^{\prime} l_p^2\,k \qquad and \qquad
\omega=\frac{c}{\bar{\lambda}}.
\end{equation}
In this situation  the dispersion relation becomes,
\begin{equation}
\omega=\omega(k)=\frac{kc}{1+\alpha^{\prime} l_p^2\,k^2}.
\end{equation}
This relation can be described in another viewpoint. By expansion of
$\Big(1+\alpha^{\prime} l_p^2\,k^2\Big)^{-1}$ and neglecting second
and higher order terms of $\alpha^{\prime}$, we find that
$\omega=kc\big(1-\alpha^{\prime} l_p^2\,k^2\big)$. This can be
considered as $\omega=kc^{\prime}$ where
$c^{\prime}=c\big(1-\alpha^{\prime} l_p^2\,k^2\big)$. This relation
indicates the possibility of variation in $c$. Accepting the
possibility of variation in $c$, one can consider its time
variation also. So we consider the time dependence of light speed.\\
Actually, there are some evidences indicating that fine structure
constant, $\alpha=\frac{e^2}{\hbar c}$ is not constant [11]. The
question then arises that which of the quantities: $e$, $c$ or
$\hbar$ are variable? A possible situation is the variation in $c$.
This is referred as varying speed of light (VSL) theories in
literatures. After introduction of this idea several varying speed
of light models have been proposed to solve the problems of standard
cosmology. One of the simplest of these models is the model proposed
by Barrow [35]. Barrow has considered the speed of light as
\begin{equation}
c(t) = c_{0}a^{n}(t),
\end{equation}
where $c_{0}$ and $n$ are constant. Using this form of $c(t)$ to
solve problems of standard cosmology, some constraint will be
imposed on the value of $n$, depending on the nature of the
problems. For example if we consider the equation of state for
matter content of the Universe as $p=(\gamma-1)\rho c^{2}(t)$,
then exact solutions with varying $c(t)$ and $G(t)$, restrict $n$
to the following limit
\begin{equation}
n \leq -1 \,\,\,\, for\,\,\, \gamma= 4/3 \quad\quad\quad Radiation
\,\, Dominated \,\, Era
\end{equation}
\begin{equation}
n \leq -1/2 \,\,\, for\,\,\, \gamma= 1 \quad\quad Matter\,\,
Dominated\,\, Era,\,Dust.
\end{equation}

\section{Black Holes Thermodynamics}
In the current standard viewpoint, small black holes emit black
body radiation at the Hawking temperature,
\begin{equation}
T_{H}\approx \frac{\hbar c^3}{8\pi GM} = \frac{M^{2}_{P}c^2}{8\pi
M},
\end{equation}
where $M_{P} =\sqrt{\frac{\hbar c}{G}}$ is the Planck mass and we
have set $k_{B} =1$. The related entropy is obtained by integration
of $dS=c^2T^{-1}dM$ which is the standard Bekenstein entropy,
\begin{equation}
S_{B}=\frac{4\pi GM^2}{\hbar c}= 4\pi \frac{M^2}{M^{2}_{P}}.
\end{equation}
If one consider the GUP as equation (2), the last two equations
become respectively,
\begin{equation}
T_{GUP}= \frac{M c^2}{4\pi}\Bigg[1\mp \sqrt{1-
\frac{M^{2}_{P}}{M^2}}\,\Bigg],
\end{equation}
and
\begin{equation}
S_{GUP}=
2\pi\Bigg[\frac{M^2}{M^{2}_{P}}\Bigg(1-\frac{M^{2}_{P}}{M^2}+\sqrt{1-
\frac{M^{2}_{P}}{M^2}}\,\Bigg) - \ln\Bigg(\frac{M+\sqrt{M^2 -
M^{2}_{P}}}{M_{P}}\,\Bigg)\,\Bigg].
\end{equation}
In equation (10), to recover the corresponding result in the limit
of large mass ($T_{H}$), one should consider the minus sign. These
equations strongly suggest the existence of black holes remnants. As
it is evident from figure 2, in the framework of GUP black hole can
evaporate until when it reachs the Planck mass. In this view point
black hole remnants are stable. Now consider the case of VSL. For
simplicity we assume that only $c$ is varying and $G$ and $\hbar$
are constant which we set $G=\hbar=1$. In this situation $
M_{P}=\sqrt{c(t)}$. This is a novel concept: a time-varying Planck
mass!. It means that Planck scales are varying with time and are
actually cosmological models dependent via dependence of $c(t)$ to
cosmological scale factor. The corresponding equations both in
Hawking-Bekenstein and GUP viewpoint will become as follows
respectively,
\begin{equation}
T^{(VSL)}_{H}(t)=\frac{c^{3}(t)}{8\pi M},
\end{equation}
\begin{equation}
S^{(VSL)}_{B}(t)=\frac{4\pi M^2}{c(t)},
\end{equation}
\begin{equation}
T^{(VSL)}_{GUP}(t)=\frac{Mc^{2}(t)}{4\pi}\Bigg[1-\sqrt{1-\frac{c(t)}{M^2}}\,\Bigg],
\end{equation}
and
\begin{equation}
S^{(VSL)}_{GUP}(t)=2\pi\Bigg[\frac{M^2}{c(t)}\Bigg(1-\frac{c(t)}{M^2}+\sqrt{1-\frac{c(t)}{M^2}}\,\Bigg)
-\ln{\Bigg(\frac{M+\sqrt{M^2-c(t)}}{\sqrt{c(t)}}\,\Bigg)}\,\Bigg].
\end{equation}
Now one should specify the time dependence of $c(t)$. Using equation
(5), since there exists several possibilities for $a(t)$ and
$c(t)=M^{2}_{P}(t)=a^{n}(t)$, we can consider de Sitter Universe as
an example,
\begin{equation}
c(t)=\Big[a_{0}\cosh(\frac{t}{a_{0}})\Big]^{n}.
\end{equation}
Now, equations (12)-(15) for de Sitter Universe become respectively,
\begin{equation}
T^{(VSL)}_{H}(t)=\frac{\cosh^{3n}(t)}{8\pi M},
\end{equation}
\begin{equation}
S^{(VSL)}_{B}(t)=\frac{4\pi M^2}{\cosh^{n}(t)},
\end{equation}
\begin{equation}
T^{(VSL)}_{GUP}(t)=\frac{M\cosh^{2n}(t)
}{4\pi}\Bigg[1-\sqrt{1-\frac{\cosh^{n}(t)}{M^2}}\,\Bigg],
\end{equation}
and
\begin{equation}
S^{(VSL)}_{GUP}(t)=2\pi\Bigg[\frac{M^2}{\cosh^{n}(t)
}\Bigg(1-\frac{\cosh^{n}(t)}{M^2}+\sqrt{1-\frac{\cosh^{n}(t)
}{M^2}}\,\Bigg)-\ln{\Bigg(\frac{M+\sqrt{M^2-\cosh^{n}(t)
}}{\sqrt{\cosh^{n}(t)}}\,\Bigg)}\,\Bigg].
\end{equation}
Where we have set $a_{0}=1$. The results of numerical calculations
are shown in figures. In these figures we have considered $n=-1$ and
the results are shown for de Sitter Universe. Note that for
different $n$, the overall behavior of the
solutions do not change considerably and other model Universes give similar results.\\
\section{Conclusions and Discussion}
Based on our model and numerical calculations, the following results are obtained \\
\begin{enumerate}
\item[1-] When one considers the time variation of speed of light alone,
total evaporation of black hole is possible in principle. Thus in
the framework of VSL micro black hole can evaporate completely. This
is in agreement with the Hawking-Bekenstein and in contrast with the
results of GUP.
\item[2-] Application of generalized uncertainty principle to black
holes thermodynamics strongly suggests the possible existence of
black holes remnant (figure 1 and 2).
\item[3-] When one considers thermodynamics of black holes in the
framework of both GUP and VSL, some novel results are obtained. The
figure for temperature of black hole versus the mass and time in a
combination of GUP and VSL (figure 5) shows that where the mass of
black hole is zero, its temperature is zero also. This is in
contrast to Hawking result and seems completely reasonable since in
the absence of matter there is no meaning for temperature. When the
mass increases, the temperature is increases until the mass becomes
equal to the Planck mass. After that, increasing of mass is
corresponding to decreasing of temperature in complete agreement
with the results of GUP (Adler {\it et al} [29]). But the situation
for mass less than Planck mass is completely different from GUP
results.
\item[4-] The figure for entropy of black hole versus the mass and time
in a combination of GUP and VSL (figure 6) shows that when one
approaches the Planck mass, entropy do not vanishes. This is
physically reasonable but rules out the result of Adler {\it et al}
since they have zero entropy for remnants. Increasing the time will
increase the entropy which is natural. The possibility of having
black hole remnant at Planck mass is evident from this figure. In
our model there is a remnant entropy for black remnant. This can be
at least related to background spacetime metric fluctuation.
\item[5-] Adler and his coworkers have constructed their formulation
based on analogy between hydrogen atom and black holes. They have
argued that since uncertainty principle prevents the hydrogen atom
from total collapse, generalized uncertainty principle may prevent
black holes total evaporation in the same manner. In our opinion,
the basic mistake of Adler {\it et al} is that they have not
considered hydrogen atom in GUP. Our calculation shows that in GUP
hydrogen atom is not stable. Since,
\begin{equation}
\Delta r\Delta p\geq \frac{\hbar}{2}(1 +\beta (\Delta p)^2).
\end{equation}
Suppose that $\Delta p\sim p$ and $\Delta r\sim r$, then one finds
\begin{equation}
pr = \frac{\hbar}{2}(1 +\beta p^2)\Rightarrow \hbar \beta p^{2} -2pr
+\hbar=0.
\end{equation}
So one obtains,
\begin{equation}
p=\frac{r\pm\sqrt{r^{2}-\beta\hbar^{2}}}{\beta\hbar}.
\end{equation}
As has been argued we should consider the minus sign in (23). Since
\begin{equation}
E = \frac{p^2}{2m} - \frac{e^2}{r},
\end{equation}
 one find,
 \begin{equation}
E=\frac{1}{2m}\bigg(\frac{r-\sqrt{r^{2}-\beta\hbar^{2}}}{\beta\hbar}\,\bigg)^2
-\frac{e^2}{r}\,.
\end{equation}
Now we set, $\frac{dE}{dr}=0$ and find
$r_{min}=\hbar\sqrt{\frac{\beta}{2}}$. The extremum value of
energy becomes,
\begin{equation}
E_{min}=-\bigg[\Big(\frac{e^2}{\hbar}\,\sqrt{\frac{2}{\beta}}\,\Big)+i\frac{1}{2m\beta}\bigg],
\end{equation}
where its real part is,
\begin{equation}
E_{min}^{real}=-\Big(\frac{e^2}{\hbar}\,\sqrt{\frac{2}{\beta}}\,\Big).
\end{equation}
Since $r_{min}$ is very small length, the radius of stability for
hydrogen atom is very small. Therefore in GUP scale, the hydrogen
atom is not stable and will collapse completely. If this is the
case, one can not construct analogy between hydrogen atom in
Heisenberg uncertainty principle viewpoint and black hole in GUP
viewpoint. If we consider hydrogen atom in GUP, as we have shown,
this atom will collapse totally. But in the framework of GUP black
holes evaporate until they reach Planck mass. The issue of stability
for remnant can be considered in the framework of symmetry principle
in the system. In this regard supersymmetry, in particular
supergravity, stands a very good
framework of providing such black hole remnants [30].\\
Note that our arguments for the existence of black hole remnants
based on GUP is heuristic. The search for its deeper theoretical
foundation is currently underway. As interactions with black hole
remnants are purely gravitational, the cross section is extremely
small, and direct observation of these remnants seems unlikely. One
possible indirect signature may be associated with the cosmic
gravitational wave background. Unlike photons, the gravitons
radiated during evaporation would be instantly frozen. Since,
according to our notion, the black hole evaporation would terminate
when it reduces to a remnants, the graviton spectrum should have a
cutoff at Planck mass. Such a cutoff would have by now been
red-shifted to $\sim10^{14} GeV$. Another possible gravitational
wave-related signature may be the gravitational wave released during
the gravitational collapse. The frequencies of such gravitational
waves would by now be in the range of $\sim 10^{7} - 10^{8} Hz$. It
would be interesting to investigate whether these signals are in
principle observable. Another possible signature may be some
imprints on the CMB fluctuations due to the thermodynamics of black
hole remnants-CMB interactions. Possible production of such remnants
in LHC (Large Hadron Collider) and also in cosmic ray showers are
under investigation. If we consider hybrid inflation as our primary
cosmological model, there will be some observational constraints on
hybrid inflation parameters. For example a simple calculation based
on hybrid inflation suggests that the time it took for black holes
to reduce to remnants is about $10^{-10} Sec$. Thus primordial black
holes have been produced before baryogenesis and subsequent epochs
in the standard cosmology [30].
\end{enumerate}
{\bf Acknowledgment}\\
This Work has been supported partially by Research Institute for
Astronomy and Astrophysics of Maragha, Iran. Also we would like to
appreciate an unknown referee of MPLA for his (her) valuable
comments on original version of the paper.\\

\begin{figure}[ht]
\begin{center}
\includegraphics{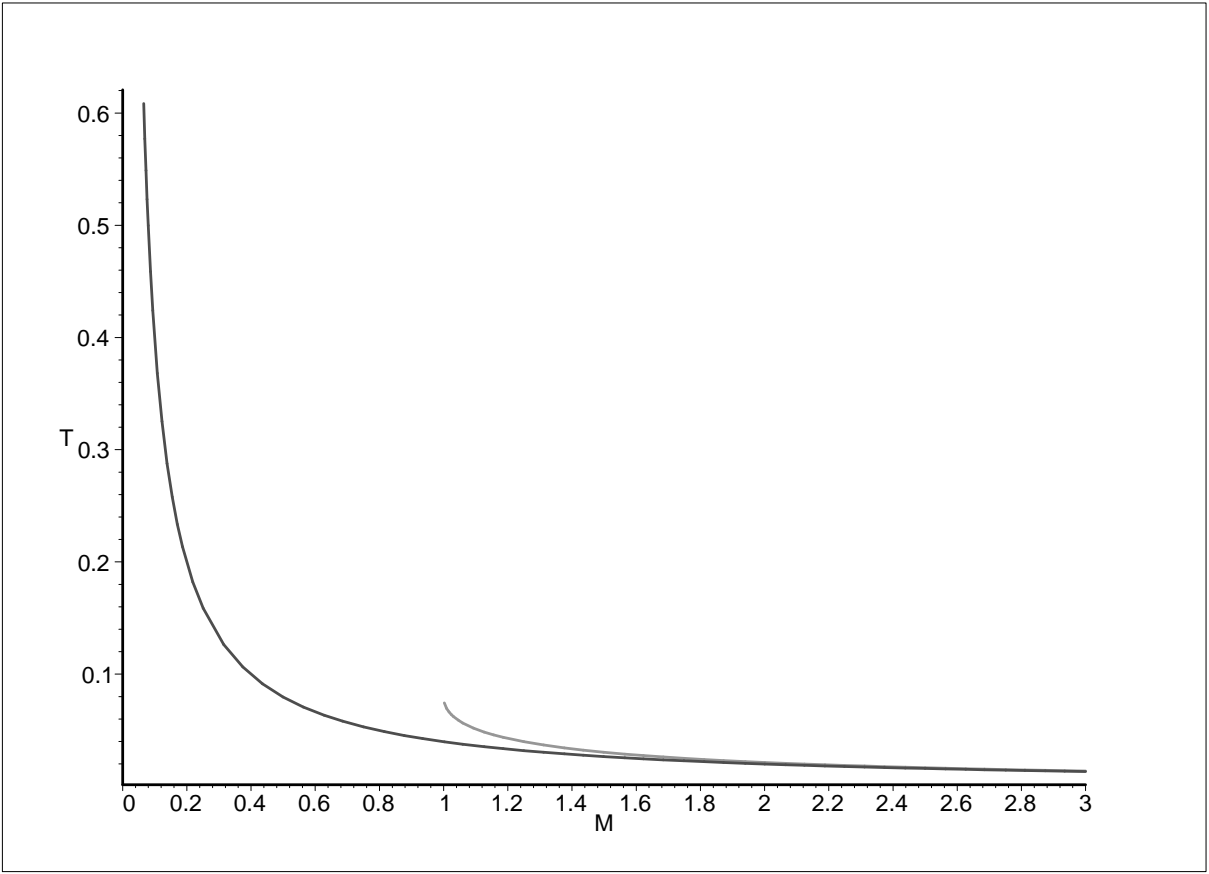}
\end{center}
\vspace{14 cm}
 \caption{\small {Temperature of a Black Hole Versus the Mass. Mass is in units of the
 Planck mass and temperature is in units of the Planck energy. The lower curve is the Hawking result,
 and the upper curve is the result of GUP. }}
\end{figure}

\begin{figure}[ht]
\begin{center}
\includegraphics{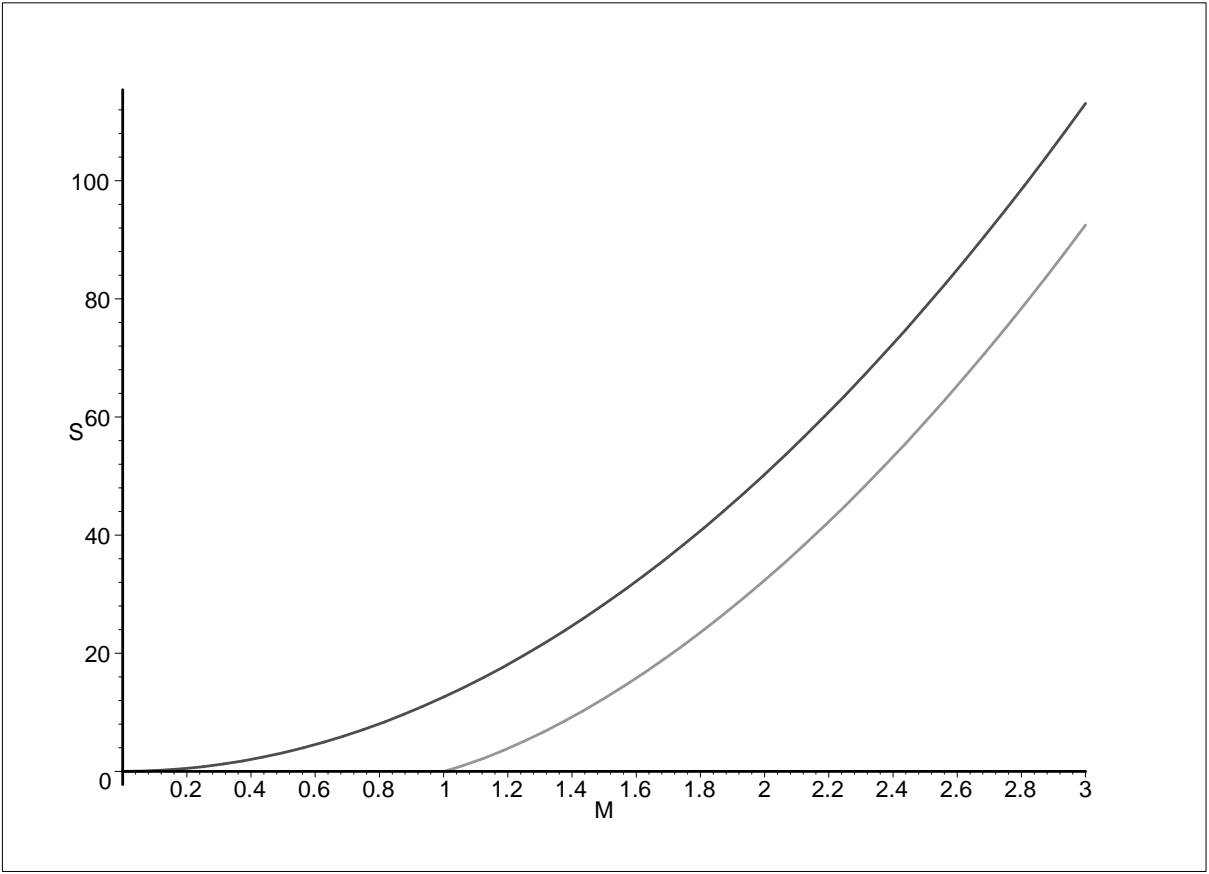}
\end{center}
\vspace{14 cm}
 \caption{\small {Entropy of a Black Hole Versus the Mass. Entropy is dimensionless
 and mass is in units of the Planck mass. The upper curve is the Hawking result, and the lower
 curve is the result of GUP. }}
\end{figure}

\begin{center}
\begin{figure}[ht]
\includegraphics{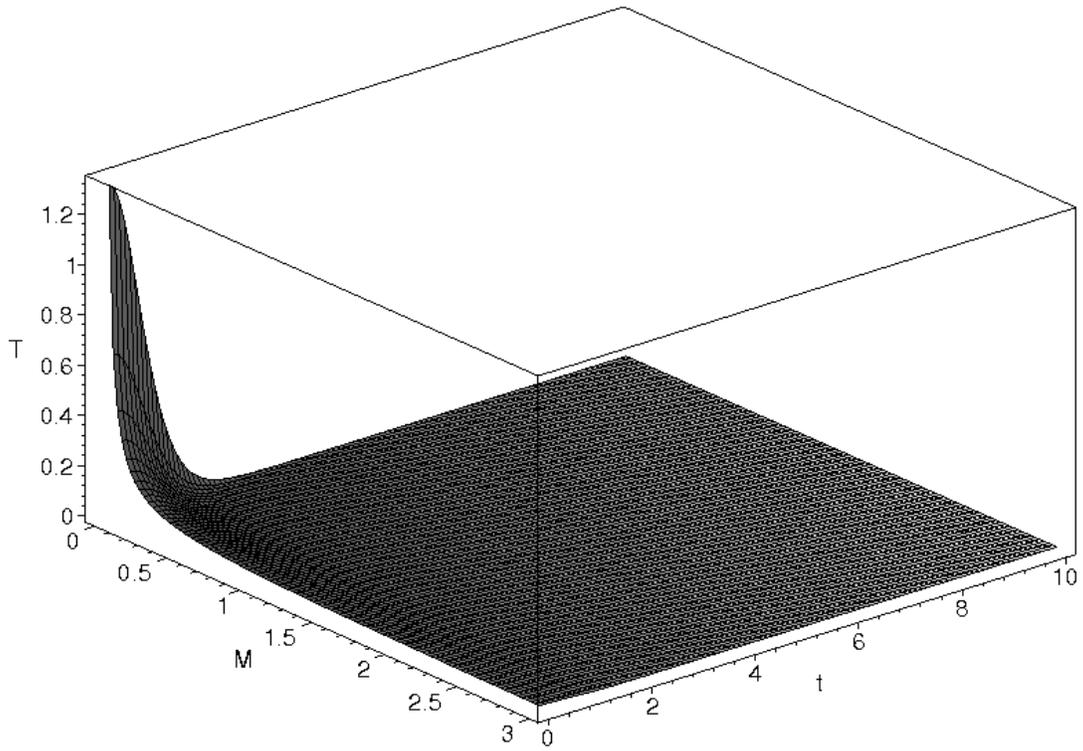} \vspace{14
cm}
 \caption{\small {Temperature of Black Hole Versus the Mass and Time in VSL.
  The units are as previous and the result is shown for
 De Sitter model. As figure shows, the result of Hawking is recovered.
 Decreasing of temperature with increasing of time is natural. }}
\end{figure}
\end{center}

\begin{center}
\begin{figure}[ht]
\includegraphics{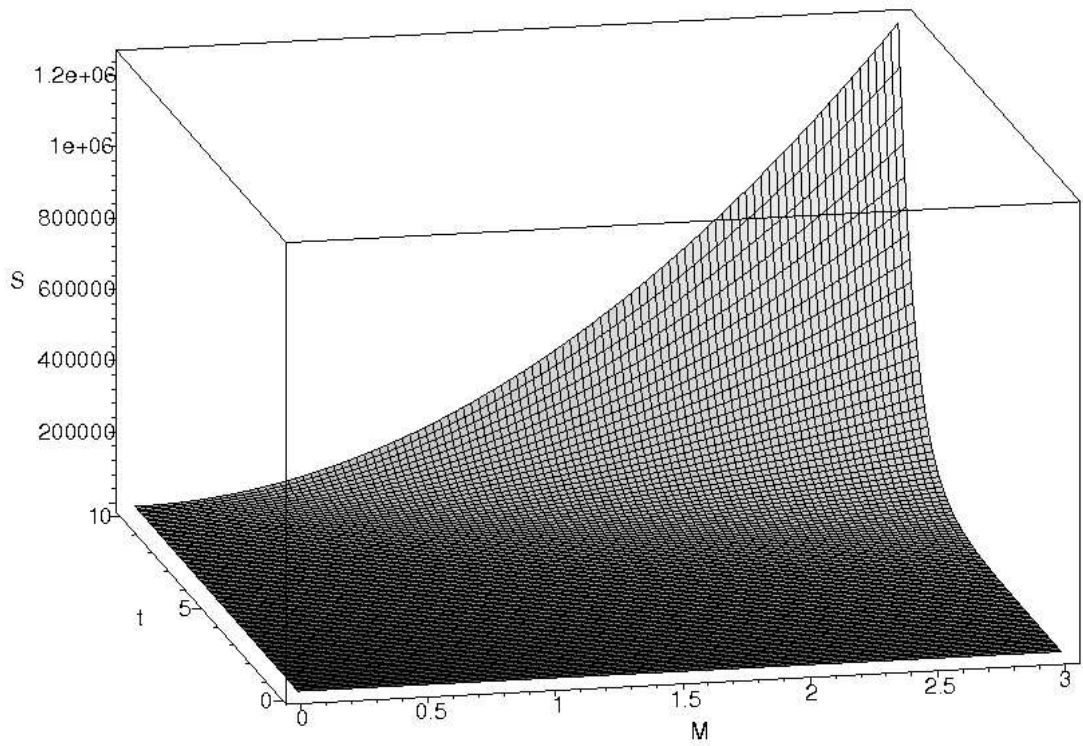} \vspace{14
cm}
 \caption{\small {Entropy of Black Hole Versus the Mass and Time in VSL.
  The units are as previous and the result is shown for
 De Sitter model. As figure shows, the result of Hawking is recovered.
 Increasing of Entropy with increasing of time is natural.}}
\end{figure}
\end{center}

\begin{center}
\begin{figure}[ht]
\includegraphics{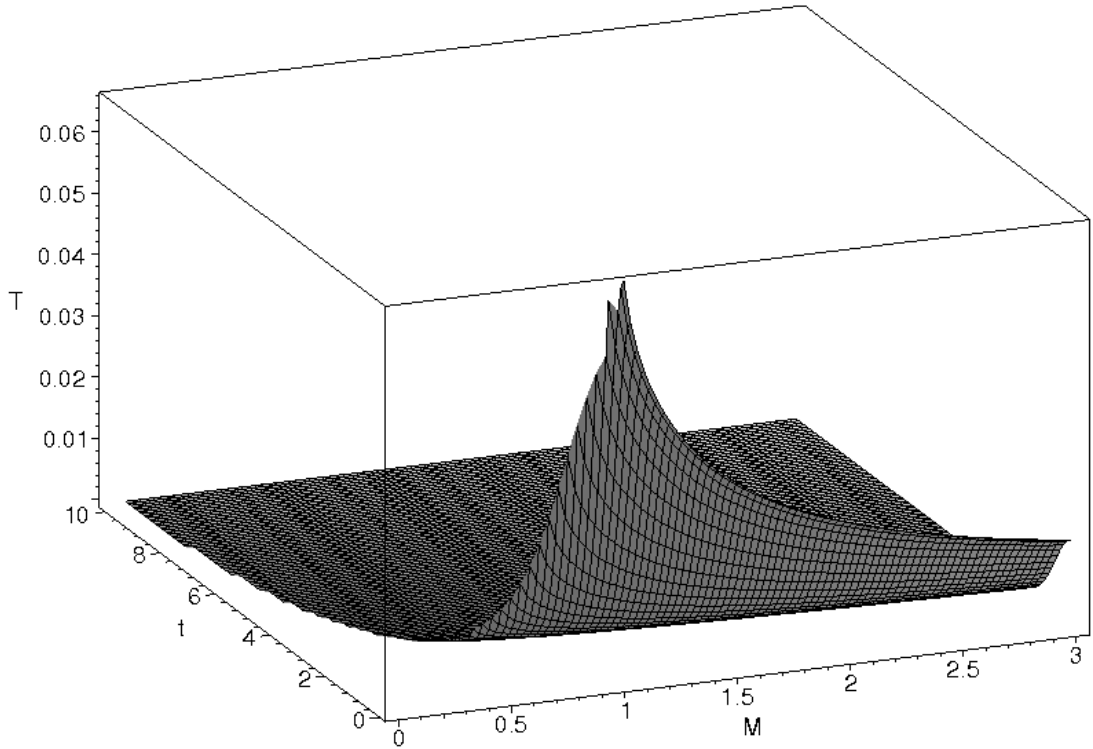} \vspace{14
cm}
 \caption{\small {Temperature of Black Hole Versus the Mass and Time in a combination of GUP and VSL.
  The units are as previous and the result is shown for
 De Sitter model. As figure shows, where the mass is zero the
 temperature is zero also. This is in contrast to Hawking result. When the mass increases,
 the temperature increases until the mass
 becomes equal to the Planck mass. After that, increasing of mass is
 corresponding to decreasing of temperature. This is a novel result of GUP+VSL Scenario.}}
\end{figure}
\end{center}

\begin{center}
\begin{figure}[ht]
\includegraphics{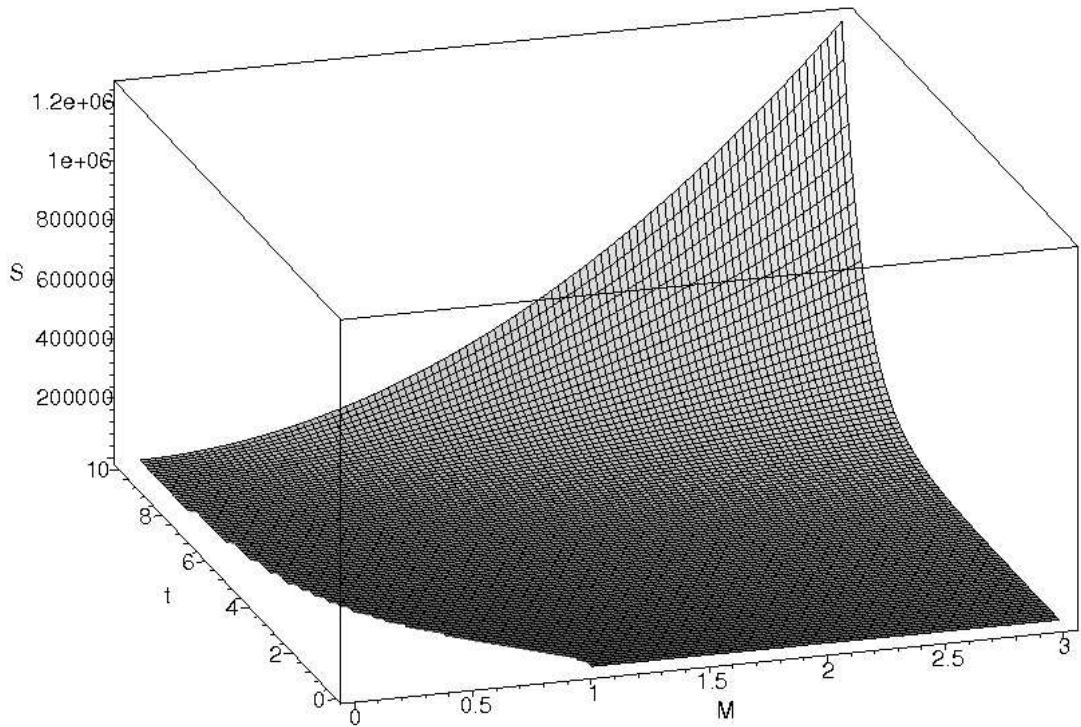} \vspace{14
cm}
 \caption{\small {Entropy of Black Hole Versus the Mass and Time in a Combination of GUP and VSL.
 The units are as previous and the result is shown for
 De Sitter model. As figure shows, When one approaches the Planck mass, entropy do not vanishes.
 Increasing entropy with time is natural.
 This figure shows the possibility of having black holes relics. }}
\end{figure}
\end{center}
\end{document}